\documentclass[a4paper]{article}

\usepackage{icrc2013}

\title{Improving the H.E.S.S. angular resolution using the Disp method}

\shorttitle{Improving the H.E.S.S. angular resolution using the Disp method}

\authors{
Chia-Chun Lu$^{1}$ for the H.E.S.S. Collaboration.
}

\afiliations{
$^1$ Max-Planck-Institut fuer Kernphysik, Saupfercheckweg 1, 69117 Heidelberg, Germany \\

}

\email{c.lu@mpi-hd.mpg.de}

\abstract{The angular resolution of imaging atmospheric Cherenkov telescopes depends on the employed event reconstruction methods. By taking the weighted average of intersections of shower axes, the H.E.S.S. experiment achieves a $0.08$ degree angular resolution at $20$ degree zenith angle with an image size cut of $160$ p.e. for sources with a spectral index of $2$. However, the angular resolution degrades to $0.14$ degree at $60$ degree zenith angle, due to the larger fraction of nearly parallel images. The Disp method reduces the impact of parallel images by including an estimation of the image displacement (disp), inferred from the Hillas parameters, in the reconstruction procedure. By using this technique, the angular resolution at large zenith angles can be improved by $50\%$. An additional cut on the estimated direction uncertainty can further improve the angular resolution to around $0.05$ degrees at the expense of a loss of $50\%$ of effective area. The performance of this 
reconstruction method on 
simulated $\gamma$-ray events 
and real data is presented.}


\begin{document}
\maketitle

\section{Introduction}
The imaging atmospheric \v Cerenkov telescopes (IACT) technique has been developed for several decades. IACTs detect the \v Cerenkov light emitted by relativistic charged particles in atmospheric air showers induced by very-high-energy ($\sim \rm{TeV}$) $\gamma$ rays. The \v Cerenkov light distribution is imaged onto the cameras. Stereoscopic observation of air showers with IACT arrays has been proven successful in providing better direction reconstruction and background rejection. The quality of direction reconstruction can be quantified by the $68\%$ containment radius, $R_{68}$, of the point-spread function which is also referred to as the angular resolution. For $\gamma$-ray sources with a given spectrum, the angular resolution depends on the design of the instruments, the analysis cuts, and the event reconstruction algorithm. As discussed in \cite{bib:Hofmann_1999}, the reconstruction algorithms mainly follow two approaches. The first kind is based on parameterizing the image as an 
ellipse, characterized by 
the first and second moments of image intensity distribution (the so-called \emph{Hillas parameters} \cite{bib:Hillas_1985}), and the second kind is based on a global fit to pixel amplitudes.

\begin{figure}[t]
  \centering
  \includegraphics[width=0.4\textwidth]{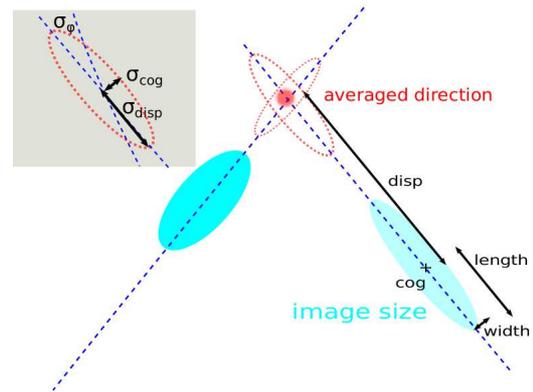}
  \caption{Schematic diagram representing the \emph{Hillas parameters} and \emph{image parameters}. The working principle of the \emph{Disp method} (\emph{Algorithm 3}) is illustrated. The definition of $\sigma_{\rm{disp}}$, $\sigma_{\rm{cog}}$, and $\sigma_{\rm{\phi}}$ are displayed in the gray box.}
  \label{fig:hillas_par}
 \end{figure}

In the analysis, the raw data is first calibrated. In the Hillas-type approach, the pixel noise produced mainly by night sky background photon is reduced by image cleaning before image parameterization. The total image intensity is denoted as \emph{size} and the center of gravity of pixel intensity distribution (the first moment) is \emph{c.o.g.} (see Figure \ref{fig:hillas_par}). The second moments of the \emph{Hillas ellipse} are the \emph{length} and \emph{width}. The orientation of major axis with respect to the x-axis of the coordinate system is defined as \emph{$\phi$}. The event direction can be calculated by \emph{Algorithm 1} in \cite{bib:Hofmann_1999} using the pair-wise intersections of extended major axes averaged by weighting factors, composed of combinations of \emph{Hillas parameters}. 

The drawback of this method is that the angular resolution degrades rapidly at larger zenith angles where the \emph{impact parameters}, defined as the perpendicular distance between the shower axis and the telescope, get on average larger. The images of showers at larger distances from the telescopes are more elongated. The axes of different images get more parallel than those of showers at smaller distances. This reflects the smaller difference in the viewing angles of different telescopes. The advantages of stereoscopic observation are thus reduced. 

\section{Method}
Event reconstruction can be improved by introducing additional \emph{image parameters}. The \emph{disp}, defined as the angular distance between \emph{c.o.g.} and event direction, the uncertainties of \emph{disp}, \emph{c.o.g.} and \emph{$\phi$}: $\sigma_{\rm{disp}}$, $\sigma_{\rm{cog}}$, and $\sigma_{\rm{\phi}}$. The prototype of this reconstruction method is referred to as \emph{Algorithm 3} in \cite{bib:Hofmann_1999}, also called the \emph{Disp method}, and used in IACT experiments such as HEGRA \cite{bib:Hofmann_1999} and VERITAS \cite{bib:Senturk_2011} for stereoscopic reconstruction. 
The \emph{image parameters} can be calculated by experiential formulae or lookup tables. In this work, multi-dimensional lookup tables filled with Monte-Carlo simulated $\gamma$-ray events are used. 
The lookup tables are two-dimensional histograms stored as a function of azimuth angle, zenith angle, \emph{impact parameter}, and optical collection efficiency. The layouts of the lookup tables are presented in Table \ref{tab:lk}. All parameters except the optical efficiency are pre-determinated by using \emph{Algorithm 1} before used in \emph{Algorithm 3} to look up the values of \emph{image parameters}. The \emph{disp} lookup tables have one extra parameter, the \emph{impact parameter}, to cope with the effect of image truncation due to the fixed $16\,$ns read-out window of the H.E.S.S. cameras. When the spread of arrival times of photons emitted in the shower on the focal plane of a camera is larger than $16\,$ns, the part of the shower beyond the read-out window is not recorded. This effect results in a reduced \emph{length} and \emph{image size} of the recorded shower image. The truncation usually happens for images of energetic showers with larger \emph{impact parameters} whose longitudinal development is more extended in time.

For each image of an event, the \emph{disp} and the orientation of the major axis yields an estimated event direction associated with an error ellipse, derived from the combination of $\sigma_{\rm{disp}}$, $\sigma_{\rm{cog}}$, $\sigma_{\rm{\phi}}$ and represented by a covariance matrix. The averaged event direction is calculated by averaging telescope-wise event directions, weighted by the inverse of the determinant of the corresponding covariance matrix. The square root of the quadratic sum of the eigenvalues of the covariance matrix associated with the final event direction is an estimate of the uncertainty on the event direction. The \emph{impact position}, which is defined as the intersection of the shower axis and the ground, is reconstructed in a similar way using \emph{impact parameters} calculated from the lookup tables.
The \emph{impact parameter} is in the first order proportional to the product of \emph{disp} and the height of shower maximum which are pre-estimated by \emph{Algorithm 1} before used to look up the new \emph{impact parameter} used in \emph{Algorithm 3}.
\begin{table}
\centering
\begin{tabular}[\textwidth]{llll} 
{\bf Output} & {\bf Parameters} & {\bf X-axis} & {\bf Y-axis} \\ 
\hline 
disp & opt,azm,zen,$\mathrm{imp}$ & $\ln$(size/p.e.) & length \\
$\mathrm{\sigma_{disp}}$ & opt,azm,zen,$\mathrm{imp}$ & $\ln$(size/p.e.) & length \\
$\mathrm{\sigma_{cog}}$ & opt,azm,zen & $\ln$(size/p.e.) & width  \\
$\sigma_{\varphi}$ & opt,azm,zen & $\ln$(size/p.e.) & width/length  \\
$\mathrm{imp'}$ & opt,azm,zen & $\mathrm{disp_{true}}$ & $\mathrm{\hat{H}_{\rm{max}}}$ \\
\end{tabular}
\caption{The layout of the image parameter (output) lookup tables used in the \emph{Disp method}. Here opt, azm, and zen are the abbreviation of telescope optical efficiency, azimuth and zenith angle of the event. The $disp$ and $\sigma_{disp}$ lookup tables have one extra parameter, $imp$, which is the \emph{impact parameter} reconstructed by \emph{Algorithm 1}. $Imp'$ is used to reconstruct the new \emph{impact position}. $\hat{H}_{\rm{max}}$ is the ratio of \emph{impact parameter} over $disp$ reconstructed by \emph{Algorithm 1}.}
\label{tab:lk} 
\end{table}

\section{Performance}
\subsection{Simulated $\gamma$-ray sources}

\begin{figure}[t]
  \centering
  \includegraphics[width=0.45\textwidth]{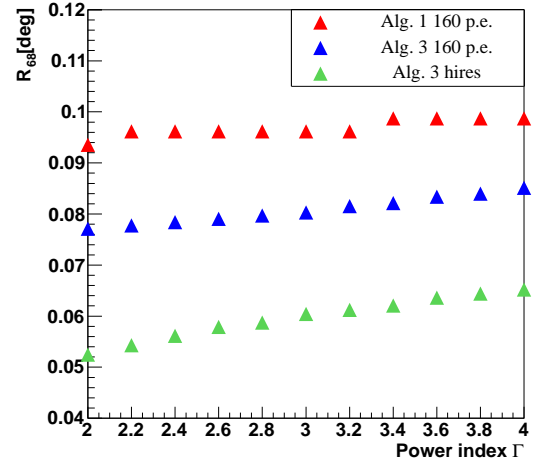}
  \caption{Angular resolution ($R_{68}$) as a function of source spectral index at $45^{\circ}$ zenith angle and $0.5^{\circ}$ offset angle for \emph{Algorithm 1} and \emph{Algorithm 3}. The angular resolution is calculated with $\gamma$-ray spectra following a power law of a spectral index $\Gamma$. The events are reconstructed with images selected with \emph{image size} larger than $160$ p.e.. The \emph{hires} configuration has an additional cut on direction uncertainty.}
  \label{fig:CTR_E}
 \end{figure}

The performance of this reconstruction method is tested with Monte-Carlo simulated $\gamma$-ray point sources under different observation conditions, e.g. different zenith angles and offset angles between the pointing direction and source direction. As shown in Figure \ref{fig:CTR_E}, $R_{68}$ is dependent on the spectral index of sources. 
Sources with softer spectra have in general poorer angular resolutions.

\begin{figure}[t]
  \centering
  \includegraphics[width=0.45\textwidth]{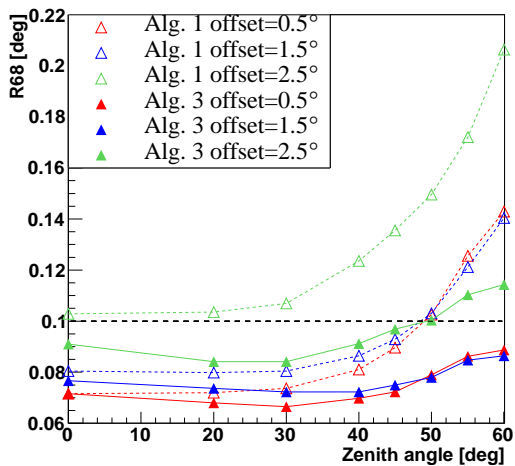}
  \caption{Angular resolution ($R_{68}$) as a function of zenith angle at different offsets for \emph{Algorithm 1} and \emph{Algorithm 3}. The events are reconstructed with images selected with \emph{image size} larger than $160$ p.e.. The dashed line denotes the radius of the integration region used for deriving the source statistics listed in Table \ref{tab:src_stat}.}
  \label{fig:CTR_zen}
 \end{figure}

$R_{68}$ of \emph{Algorithm 1} and $3$ as a function of zenith and offset angles is presented in Figure \ref{fig:CTR_zen}. The $\gamma$-ray energy distribution is assumed following a power law with a spectral index of $2$. One can see that $R_{68}$ of \emph{Algorithm 3} degrades more slowly with increasing zenith and offset angles than that of \emph{Algorithm 1}. For an offset of $2.5^{\circ}$, the $R_{68}$ is significantly improved by \emph{Algorithm 3} because of the increased fraction of showers at large distances from the telescopes. 

\begin{figure}[t]
  \centering
  \includegraphics[width=0.45\textwidth]{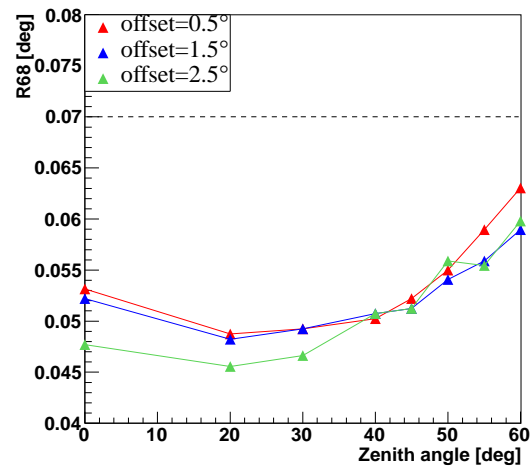}
  \caption{Angular resolution ($R_{68}$) as a function of zenith angle at different offsets for the \emph{hires} configuration. The events are reconstructed with images selected with \emph{image size} larger than $160$ p.e.. An extra cut on direction uncertainty of smaller than $0.08^{\circ}$ is applied after reconstruction. The dashed line denotes the radius of the integration region used for deriving the source statistics listed in Table \ref{tab:src_stat}.}
  \label{fig:CTR_zen_hires}
 \end{figure}

\begin{figure}[t]
  \centering
  \includegraphics[width=0.45\textwidth]{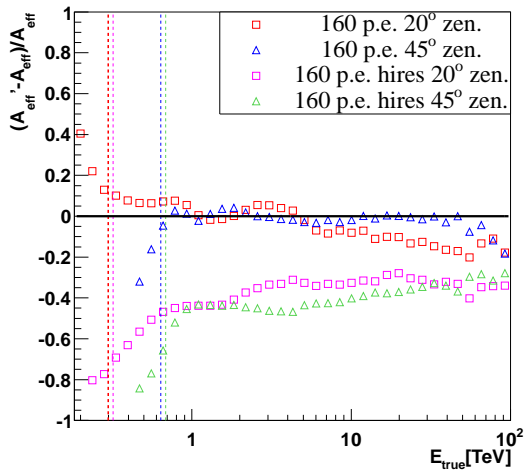}
  \caption{Relative difference in effective area as a function of energy at $20^\circ$ and $45^\circ$ zenith angles. $A_{\rm{eff}}'$: Effective area obtained by using \emph{Algorithm 3}. $A_{\rm{eff}}$: By using \emph{Algorithm 1}. Dashed lines denote the safe energy threshold above which the energy bias is smaller than $10\%$.}
  \label{fig:EffA_Diff1}
 \end{figure}

\begin{figure}[t]
  \centering
  \includegraphics[width=0.45\textwidth]{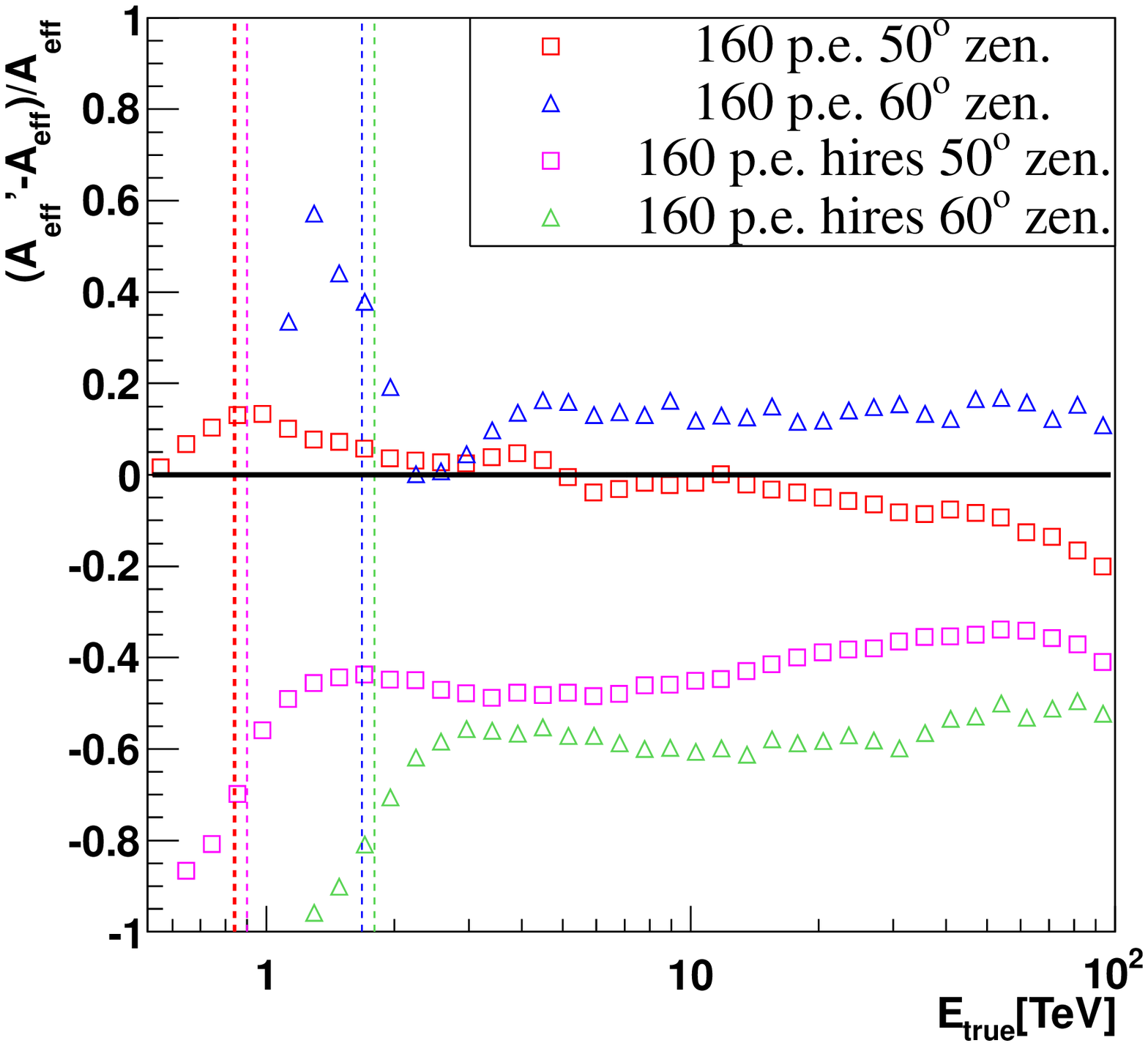}
  \caption{Relative difference in effective area as a function of energy at $50^\circ$ and $60^\circ$ zenith angles.}
  \label{fig:EffA_Diff2}
 \end{figure}

Figures \ref{fig:EffA_Diff1} and \ref{fig:EffA_Diff2} shows the change of effective area by using \emph{Algorithm 3}. Mainly two factors contribute to the change: The changes of the angular resolution and of the $\gamma$-hadron separation cut. At $60^{\circ}$ zenith angle, the effective area is significantly increased due to the much better angular resolution. By introducing an extra cut of $0.08^{\circ}$ on the direction uncertainty, $R_{68}$ can be reduced to $\sim 0.05^{\circ}$ as shown in Figure \ref{fig:CTR_zen_hires} at the expense of a loss of $\sim 50\%$ of effective area. The effective area of low energy $\gamma$-rays by this so-called \emph{hires} configuration is significantly reduced as shown in Figures \ref{fig:EffA_Diff1} and \ref{fig:EffA_Diff2} so it is more suitable for the analysis of sources with hard spectra. 
\subsection{Real data}
Two well known point-like sources are used to compare the performance of the two presented methods: Mkn 421 and the Crab nebula. Mkn 421 is a bright BL Lac object at a redshift of $z=0.031$ and was reported as a VHE $\gamma$-ray source in \cite{bib:Mkn421_HESS}.
The Crab nebula is a young pulsar wind nebula at a distance of $2\,$kpc. Given its high and steady flux at energies above $100$ GeV, the Crab nebula is commonly used as a standard candle for VHE $\gamma$-ray telescopes. The H.E.S.S. detection of the Crab nebula was published in \cite{bib:Crab_HESS}. Both these two sources have a spectrum with spectral index softer than $2$ and a cutoff at high energy.

The analysis of the two objects presented here is performed using the H.E.S.S. analysis software \emph{hap} with direction reconstruction following \emph{Algorithm 1} and $3$. A fraction of the data used in the original publication is used, resulting in a mean observation zenith angle of $61^{\circ}$ for Mkn 421 and $47^{\circ}$ for the Crab nebula. $\gamma$-like events are selected following a Multivariate Analysis technique \cite{bib:Ohm_2009}. In this technique, parameters derived from image shapes 
and reconstructed physical quantities of the shower such as the depth of shower maximum and the sample standard deviation of 
the energies reconstructed by participating telescopes are used to reject the background-like events. 

\begin{figure}[]
  \centering
  \includegraphics[width=0.45\textwidth]{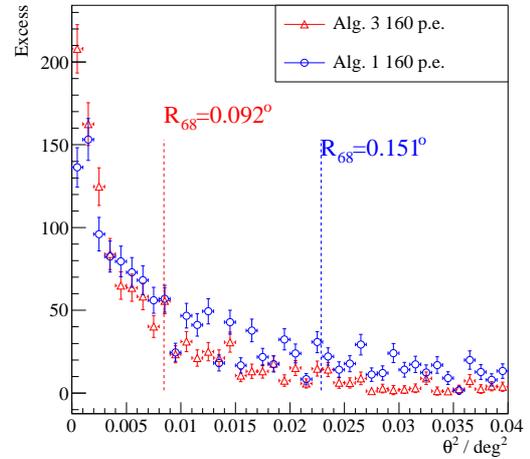}
  \caption{Squared angular ($\theta^2$) distribution of the excess events for Mkn 421. $R^2_{68}$ is denoted by the dashed line. The angular distance is calculated with respect to the position of Mkn 421.}
  \label{fig:thetasq_mkn421}
 \end{figure}

\begin{figure}[]
  \centering
  \includegraphics[width=0.45\textwidth]{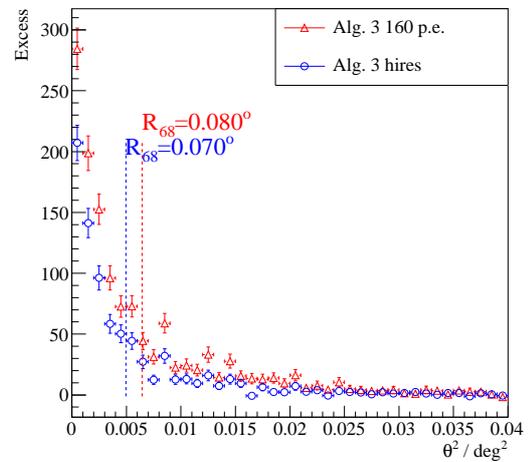}
  \caption{Squared angular ($\theta^2$) distribution of the excess events for the Crab nebula. $R^2_{68}$ is denoted by the dashed line. The angular distance is calculated with respect to the position of the Crab pulsar.}
  \label{fig:thetasq_crab}
 \end{figure}

The statistics of $\gamma$-like events are summarized in Table \ref{tab:src_stat}. The \emph{ring background} method is used to derive these quantities following the procedure described in \cite{bib:Crab_HESS}. Due to the stricter $\gamma$-hadron separation cut, the significance of Crab nebula is not changed much although the $R_{68}$ is improved by using \emph{Algorithm 3}.\footnote{The analysis cuts presented here are optimized for sources with a spectral index of 2 and $1\%$ Crab flux.}
The significance of Mkn 421 is significantly increased due to larger excess of $\gamma$-like events from the direction of the target. 
The squared angular distributions of $\gamma$-like excess events are presented in Figures \ref{fig:thetasq_mkn421} and \ref{fig:thetasq_crab}. The distribution of Mkn 421 using \emph{Algorithm 3} has a more pronounced peak around the target position and a shorter tail compared with that by \emph{Algorithm 1}. The distribution of the Crab nebula by the \emph{hires} cut has a $\sim 17\%$ smaller $R_{68}$ and also a shorter tail compared with that by \emph{Algorithm 3} with the $160$-p.e size cut. The cut on the direction uncertainty reduces the number of $\gamma$-like events by $\sim 50\%$ but rejects even more hadronic events so the detection significance of point-like sources is kept $\sim 80\%$ as good as the configuration without this cut.  
\begin{table*}
\centering
\begin{tabular}{ccccccc}
\hline
Target & Config. & $N_{\rm{On}}$ & $N_{\rm{Off}} \cdot \alpha$  & Excess & Sig.[$\sigma$] & $R_{68}$[deg] \\
\hline
Mkn 421 & Alg. 1 & 884 & 50 & 834 & 54 & 0.151\\
Mkn 421 & Alg. 3 & 914 & 23 & 891 & 63 & 0.092\\
Mkn 421 & hires  & 506 &  4 & 502 & 57 & 0.085\\
Crab    & Alg. 1 & 1175& 32 & 1143& 70 & 0.094\\
Crab    & Alg. 3 & 1056 & 21 & 1035 & 69 & 0.080\\
Crab    & hires  & 548 & 4 & 544 & 58  & 0.070\\
\hline
\end{tabular}
\caption{Statistics of $\gamma$-ray-like events from Mkn 421 and the Crab nebula. The image \emph{size} cut is 160 p.e..
 $N_{\rm{On}}$ is the event count within a circular integration region of the radius optimized for point-like source extraction which is $0.07^{\circ}$ for the hires configuration and $0.1^{\circ}$ for other configurations. $N_{\rm{Off}} \cdot \alpha$ is the normalized background count. The excess is given as $N_{\rm{On}}- \alpha N_{\rm{Off}}$ and the significance (Sig.) is calculated following the formula $17$ in \cite{bib:LiMa}.}
\label{tab:src_stat} 
\end{table*}
\section{Conclusions}
In this work, the performance of the direction reconstruction technique using the \emph{Disp method} (\emph{Algorithm 3}) is presented and compared with that of \emph{Algorithm 1}. The angular resolution by \emph{Algorithm 1} degrades rapidly with the zenith angle and is significantly improved by the \emph{Disp method} at zenith angles larger than $45^{\circ}$ . For events from large zenith angles of between $45^{\circ}$ and $60^{\circ}$ at $0.5^{\circ}$ to $1.5^{\circ}$ offsets, the improvement is $20\%-40\%$. For offsets larger than $2.0^{\circ}$, there is an additional improvement of $5\%-10\%$. 
The \emph{hires} configuration with an extra cut on the direction uncertainty achieves an angular resolution of
$\sim0.05^{\circ}$ at the expense of a loss of $50\%$ of effective area. 

This reconstruction technique can be applied widely to various kinds of sources taking the advantages of better angular resolution. The \emph{hires} configuration with a significantly improved angular resolution is especially suitable for studies of sources with complicated morphology but high event statistics. 

\vspace*{0.5cm}
\footnotesize{{\bf Acknowledgment:}{Please see standard acknowledgement in
H.E.S.S. papers, not reproduced here due to lack of space.}


\begin{thebibliography}{}
\bibitem{bib:Hofmann_1999} W. Hofmann \emph{et al.}, Astroparticle Physics 12 (1999) 135-143 doi:10.1016/S0927-6505(99)00084-5
\bibitem{bib:Hillas_1985} A. M. Hillas, Proceedings of the 19th International Cosmic Ray Conference (1985) 445-448
\bibitem{bib:Senturk_2011} G. D. Senturk, Proceedings of the 32nd International Cosmic Ray Conference (2011) 126-128
\bibitem{bib:Crab_HESS} F. Aharonian \emph{et al.}, Astronomy and Astrophysics 457 (2006) 899-915 doi:10.1051/0004-6361:20065351
\bibitem{bib:Mkn421_HESS} F. Aharonian \emph{et al.}, Astronomy and Astrophysics 437 (2005) 95-99 doi:10.1051/0004-6361:20053050

\bibitem{bib:Ohm_2009} S. Ohm \emph{et al.}, Astroparticle Physics 31 (2009) 383-391 doi:10.1016/j.astropartphys.2009.04.001
\bibitem{bib:LiMa}  T.-P. Li and Y.-Q. Ma, Astrophysical Journal 272 (1983) 317-324 doi:10.1086/161295
\end{thebibliography}
\end{document}